\documentclass[%
 reprint,
superscriptaddress,
showpacs,
 amsmath,amssymb,
 aps,
]{revtex4-1}

\usepackage{graphicx}
\usepackage{color}
\usepackage{verbatim}
\usepackage{subfigure}

\definecolor{Red}{rgb}{0,0,0}
\definecolor{Blue}{rgb}{0,0,0}
\newcommand{\NB}[1]{{\color{Blue} #1}}
\newcommand{\NBc}[1]{{\color{Red} #1}}

\begin{document}

\preprint{APS/123-QED}

\title{Coupling quasi-phase matching: \\ entanglement buildup in $\chi^{(2)}$ nonlinear waveguide arrays} 

\author{David Barral} 
\email{Corresponding author: david.barral@c2n.upsaclay.fr}
\author{Nadia Belabas}
\author{Kamel Bencheikh}
\author{Juan Ariel Levenson}
\affiliation{Centre de Nanosciences et de Nanotechnologies C2N, CNRS, Universit\'e Paris-Saclay, 10 Boulevard Thomas Gobert, 91120 Palaiseau, France}

\begin{abstract} Wavevector quasi-phase matching was devised in the sixties as a way to boost nonlinear interactions with efficient quantum noise squeezing as one outstanding outcome. In the era of quantum technologies, we propose a new coupling quasi-phase matching for efficient generation of multimode downconverted quantum light in nonlinear waveguide arrays. We highlight this technique achieving multimode quantum entanglement and Einstein-Podolsky-Rosen steering buildup. We discuss the feasibility of this method with current technology and demonstrate its competitiveness as a resource for continuous variables quantum information.
\end{abstract}

\date{June 11, 2019}
\maketitle 

\section{Introduction} The buildup of second-order nonlinear interactions is directly related to the ability to propagate the interacting waves at the same phase velocity, i.e. the phase matching condition. The most efficient strategy, birefringence, is not always sufficient to compensate the phase mismatch (e.g. semiconductors) or it is not applicable to the highest second-order tensor component (e.g. lithium niobate) \NB{for high conversion efficiency}. Quasi-phase matching (QPM) is nowadays the usual name of a clever solution introduced in the seminal paper of Armstrong et al. \cite{Armstrong1962}. This method is based on periodical reset of \NB{wavevector} phase mismatch to maintain a coherent buildup of the nonlinear interaction \NB{ and it is conventionally obtained by periodic modulation of the nonlinear coefficient \cite{Somekh1972}}. The QPM approach was successfully extended to several situations where the mismatch could be compensated at first, second or nth order, or tailored to enhance cascaded second-order nonlinearities \cite{Hum2007}. {Optical-field noise squeezing and twin photons are produced in quantum-optics labs worldwide using this technique \cite{Andersen2016}. 

The key resource of disruptive quantum technologies is entanglement and the quest of efficient multimode sources is a thriving area of research \cite{Acin2018}.} In table-top bulk-optics experiments entanglement is typically generated in nonlinear crystals either by suitable (quasi-)phase matching of nondegenerate fields or by linear interaction in beamsplitters of degenerate squeezed light \cite{Lvovsky2015}. However, these resources are far away from real-world technology: they are {neither off-the-shelf nor} compact, stable or low-cost. Integrated and fiber optics are strong candidates to take over \cite{Tanzilli2012}. Entangled states of light have been indeed produced through sequential production of squeezed light and injection in directional couplers, which couple the propagating modes through evanescent-field tails \cite{Jin2014, Lenzini2018}. {Remarkably, coupling can be incorporated differently to enable another class} of integrated-optics elements without bulk-optics analogous: nonlinear waveguide arrays which rely on distributed coupling and nonlinearity \cite{Solntsev2014, Kruse2015, Setzpfandt2016}, {i.e. light undergoes coupling and nonlinearity simultaneously and not sequentially}. The phase matching in these compact, novel and versatile devices is not trivial since a cascade phase mismatch is introduced by the evanescent coupling between waveguides with impact on the nonlinear efficiency \cite{Herec2003, Barral2017}. Some strategies have been developed to avoid this detrimental effect: intensity modulation-based QPM has been proposed for second harmonic generation in coupled waveguides \cite{Huang1998, Dong2004}, and that approach has been extended recently to sum-frequency, difference-frequency and third harmonic generation \cite{Biaggio2014, Huang2015}. {These relevant proposals however, do not take advantage of the eigenmodes --or supermodes-- of the linear array \cite{Kapon1984}, are not combined with wavevector QPM ($\Delta \beta$-QPM) and deal with classical light. In this paper we introduce coupling-QPM (C-QPM) and we show how C-QPM can be seen as phase matching of the array supermodes. Our technique can be combined with the usual wavevector phase mismatch compensating technique to achieve a continuous growth of the nonlinear interaction for certain eigenmodes of the linear system. We thus further demonstrate the powerful potential of C-QPM as a quantum resource as this continuous nonlinear interaction results in strong entanglement between the individual elements of these eigenmodes}. We focus on the simplest case, the emblematic nonlinear directional coupler (NDC), and analyze its performance in the spontaneous parametric down-conversion (SPDC) regime. We study the abilities of the C-QPM NDC in the framework of continuous-variables (CV). {Integrated CV quantum information is indeed a thriving area of research \cite{Dutt2015, Masada2015, Kaiser2016, Stefszky2017, Lenzini2018, Mondain2018} and it includes discrete variables regime as a limit case. We illustrate the impact of C-QPM on} CV quantum information features such as noise squeezing, quantum entanglement and Einstein-Podolsky-Rosen (EPR) steering, and conclude discussing the feasibility of our method.
\begin{figure}[t]
\centering
\includegraphics[width=0.49\textwidth]{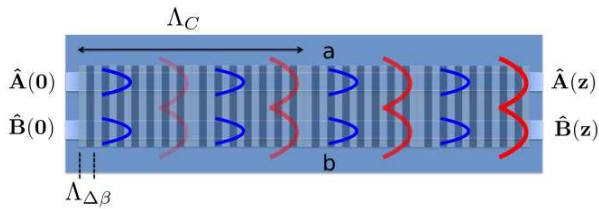}
\vspace {0cm}\,
\hspace{0cm}\caption{\label{F1}\small{(Color online) Sketch of the proposed nonlinear directional coupler made of two identical waveguides {\it a} and {\it b} with second-order susceptibilities $\chi^{(2)}$. In blue the non-interacting pump waves. In red the efficiently buildup signal waves. $\Lambda_{\Delta\beta}$ stands for the period of the uniform grating. $\Lambda_{C}$ stands for the period of the phase-reversal grating (superperiod).}}
\end {figure}


\section{Coupling Quasi-Phase Matching} The NDC, sketched in Figure \ref{F1}, is made of two identical $\chi^{(2)}$ waveguides in which degenerate SPDC takes place \cite{Perina2000}. In each waveguide, a pump photon (p) at  frequency $\omega_{p}$ is downconverted into indistinguishable idler (i) and signal (s) photons with equal frequencies $\omega_{s,i}=\omega_{p}/2$ and identical polarization modes (type-$0$ process). The efficiency of the nonlinear interaction in a single waveguide depends on the propagation constants mismatch between the pump and signal photons caused by dispersion $\Delta \beta \equiv \beta(\omega_{p})-2 \beta(\omega_{s})$, with $\beta(\omega_{s,p})$ the propagation constant corresponding to frequency $\omega_{s,p}$. 
A common implementation of $\Delta \beta$-QPM is periodical inversion of the second-order susceptibility $\chi^{(2)}$, like for instance in periodically poled lithium niobate waveguides (PPLN) \cite{Alibart2016}. The energy of the downconverted signal modes is exchanged between the waveguides through evanescent tails, resulting into a linear coupling $C$ of the signal fields, whereas the interplay of the higher frequency pumps is negligible for the considered propagation lengths. The pump can be safely assumed undepleted if strong coherent pumps are used. The production of CV entangled states in a perfectly phase-matched NDC has been theoretically studied in refs.\,\cite{Herec2003, Barral2017, Barral2018}. These works demonstrated that in the technologically available range of CW-PPLN directional couplers --where linear coupling dominating over nonlinear coupling and sample lengths are a few centimeters-- the entanglement between the two signal output fields is maximized for equal input pump powers and phases, showing an oscillatory evolution which periodically shifts between a maximum and zero values \cite{Note0}. This detrimental periodic evolution arises from a coupling-based nonlinear phase mismatch \NB{and limits the amount of available entanglement. We introduce below C-QPM as a method to avoid this unfavorable consequence and show its effect on equal input pump power and phase in each waveguide.}

\NB{The relevant operator which describes the propagation in this system is the interaction momentum, which can be written as follows \cite{Barral2017}
\begin{equation}\nonumber
\hat{M}=\hbar\, \{C \hat{A} \hat{B}^{\dag} + \eta \, e^{i(\Delta \beta z + \phi)} (\hat{A}^{\dag\,2}+\hat{B}^{\dag\,2}) + H.c.\},
\end{equation}
where $\hat{A}$ and $\hat{B}$ are slowly varying amplitude annihilation operators of signal photons in the upper ({\it a}) and lower ({\it b}) waveguides, respectively, $\eta_{a}=\eta_{b}\equiv \eta$ is the nonlinear constant proportional to $\chi^{(2)}$ and to the pump power coupled into the waveguides, $C$ the linear coupling constant, $\hbar$ the Planck constant, $\phi_{a}=\phi_{b}\equiv \phi$ is the input phase of the pump fields, $z$ is the coordinate corresponding to the direction of propagation, and ${\it H.c.}$ stands for Hermitian conjugate. $C$ has been taken as real without loss of generality. From this momentum operator, the following Heisenberg equations are obtained
\begin{align}\nonumber
\frac{d \hat{A}}{d z}=i C \hat{B} + 2 i  \eta \, e^{i(\Delta \beta z + \phi)} \hat{A}^{\dag}, \\ \label{QS0}
\frac{d \hat{B}}{d z}=i C \hat{A} + 2 i  \eta \, e^{i(\Delta \beta z + \phi)} \hat{B}^{\dag}.
\end{align}
These equations hold all the dynamical information of the system. However, the individual modes basis hides the coupling-based phase introduced above. The natural basis for this problem is indeed the eigenmodes basis --supermodes-- for the evanescently coupled signal modes \cite{Kapon1984}. The Heisenberg equations (\ref{QS0}) take in this basis the following simple form
\begin{align}\nonumber
\frac{d \hat{E}}{d z}=2 i  \eta \, e^{i(\Delta \beta z -2Cz + \phi)} \hat{E}^{\dag}, \\ \label{QS1}
\frac{d \hat{O}}{d z}=2 i  \eta \, e^{i(\Delta \beta z +2Cz + \phi)} \hat{O}^{\dag},
\end{align}
where we have introduced the even and odd supermode operators $\hat{E}$, $\hat{O}$, defined as
\begin{equation}\nonumber
\hat{E}= \frac{\hat{A}+\hat{B}}{\sqrt{2}} e^{-i\, C z },\quad
\hat{O}= \frac{\hat{B}-\hat{A}}{\sqrt{2}} e^{ i\, C z }.
\end{equation}
}These are the equations for two decoupled parametric amplifiers in the supermodes basis with {a $z$-dependent} gain. This longitudinal dependence degrades periodically the amplifier gain with fast and slow periods respectively related to the wavevector and coupling phase mismatches as $\Delta\beta$ is few orders of magnitude higher than $C$ in general. Equations (\ref{QS1}) thus suggest that a suitable tailoring of the effective nonlinearity $\eta$ through periodic domains and super-domains could compensate both phase mismatches and would lead to an efficient amplification of the supermodes. A suitable engineering of the nonlinear parameter can be then
\begin{equation}\label{etaZ}
\eta=\eta_{0} f_{\Delta\beta}(z) f_{C}(z),
\end{equation}
with $f_{\Delta\beta}(z)$ and $f_{C}(z)$ standing respectively for fast and slow periodical square-wave domains along propagation with duty cycles of 50\% (Figure \ref{F1})  \cite{Fejer1992}. By inspection of Equations (\ref{QS1}), the wavevector $\Lambda_{\Delta\beta}$ and coupling $\Lambda_{C}$ periods can be tentatively set as $\Lambda_{\Delta\beta}=2\pi/\Delta\beta$ for a uniform wavevector fast grating and $\Lambda_{C}=\pi/C \gg \Lambda_{\Delta\beta}$ for a coupling phase-reversal slow grating. The use of $\Lambda_{\Delta\beta}$ and $\Lambda_{C}$ superperiods leads to a coherent buildup of the nonlinear interaction at the cost of a reduction in the effective nonlinearity to
\begin{equation}\label{RelEta}
\eta_{C}=(2/\pi)\, \eta_{\Delta\beta}=(2/\pi)^{2}\, \eta_{0},
\end{equation}
for first order $\Delta\beta$- and C-QPM. \NB{This drop in the nonlinear efficiency can be compensated by a longer interaction distance.} Engineered phase-reversal gratings in PPLN have been reported in single waveguides to produce multiple wavelength conversion \cite{Chou1999}. In our case wavelength degeneracy is preserved by the coupling phase matching. In waveguide arrays, it has been recently shown that suitable engineering of nonuniform poling domains can produce any set of path-entangled biphoton states in the discrete-variables regime \cite{Titchener2015}. Nevertheless, this technique can be technologically demanding and prone to fabrication errors. In contrast, our approach involves two poling periods that compensate for the fast and slow mismatches. Indeed, for propagation distances $z\gg\Lambda_{C}$, Equations (\ref{QS1}) are approximated by
\begin{equation} \label{QS2}
\frac{d \hat{E}}{d z}\approx 2 i  \eta_{C} \, e^{i \phi} \hat{E}^{\dag}, \qquad \frac{d \hat{O}}{d z}\approx 2 i  \eta_{C} \, e^{i \phi} \hat{O}^{\dag}. 
\end{equation}
These equations are analogous to the evolutions of fields in individual waveguides with ideal perfect wavevector phase matching (PPM) in the undepleted pump regime \cite{Huttner1990}. This analysis explains why the proposed C-QPM is indeed QPM at the supermodes level and brings out the mechanisms at play. Notably, it shows that this approach is scalable to any number of waveguides as it relies on the supermodes only. \NB{In general, $\Lambda_{C}(k)$ will match the propagation constant $\beta_{k}^{S}$ of the $k$-th slowly varying supermode amplitude. In the case of homogeneous arrays, $\beta_{k}^{S}=-\beta_{N+1-k}^{S}\equiv 2C \cos{(k\pi/(N+1))}$ with N the number of waveguides \cite{Kapon1984}, and the coupling period can be set as 
\begin{equation}\nonumber
\Lambda_{C}({k})=\left| \frac{\pi}{2 C\cos{(k\pi/(N+1)})} \right|,
\end{equation}
thus phase matching the $k$-th and $(N+1-k)$-th supermodes. Note that in the case of odd number of waveguides N, the supermode $k=(N+1)/2$ is phase matched without the use of C-QPM since $\beta^{S}_{(N+1)/2}=0$ \cite{Barral2019}. This supermode is then the only one efficiently building up. Remarkably, C-QPM opens up the possibility to efficiently buildup any pair ($k,N+1-k$) of supermodes. The above analysis is also valid for optical parametric amplification in the classical regime. }

Therefore, numerical analysis of Equations (\ref{QS1}-\ref{etaZ}) could in principle be enough to compute optical-fields propagation, squeezing and entanglement. Nevertheless, the above analysis does not give the full picture: exactly compensating phase for mismatch involves a nonlinear dependence of the wavevector $\Lambda_{\Delta\beta}$ and coupling $\Lambda_{C}$ periods. Since $\Delta\beta \gg \eta$, $\Lambda_{\Delta\beta}=2\pi/\Delta\beta$ is a suitable selection \cite{Bencheikh1995}. However, in the case of the coupling period $\Lambda_{C}$, the nonlinearity has a stronger influence since coupling and nonlinearity can present similar orders of magnitude depending on the pump power. Typical values in PPLN waveguides are $\Delta\beta=20.10^{-2}\, \mu m^{-1}$, $C=36.10^{-2}$ mm$^{-1}$ and $\eta_{0}=15.10^{-4} \sqrt{P}$ mm$^{-1}  mW^{-1/2}$, with P the input pump power \cite{Kruse2015, Lenzini2018}. Thus a thorough analysis is needed to validate $\Lambda_{C}=\pi/C$ as a good setting. Moreover, a phase retardation $-\pi/2$ appears in $\Delta\beta$-QPM single waveguides and affects quantum noise squeezing \cite{Bencheikh1995}. Similarly, coupling-based phases can play a role in the entanglement of the fields in the case of evanescently coupled waveguides. Thus, we present here a full model of the propagation in a C-QPM NDC by studying analytically the propagation at the level of each wavevector period. This analysis gives insight to correctly set $\Lambda_{C}$ and includes and recovers all the relevant phases.


\section{Propagation in a C-QPM NDC: full model} Our full model consists in the study of the propagation at the level of each inversion period. Since we are interested in CV squeezing and entanglement of the individual fields, it is more convenient to deal with the field quadratures  $\hat{X}_{(A,B)}$, $\hat{Y}_{(A,B)}$, where $\hat{X}_{S}=(\hat{S}+\hat{S}^{\dag})/\sqrt{2}$ and $\hat{Y}_{S}=i (\hat{S}^{\dag}-\hat{S})/\sqrt{2}$ are, respectively, the orthogonal amplitude and phase quadratures corresponding to a signal optical mode $S\equiv A, B$. The system of Equations ($\ref{QS0}$) can be rewritten as $d \hat{\xi}/ d z = \mathbf{\Delta}(z)\, \hat{\xi}$ in terms of the individual-modes quadratures, where $\mathbf{\Delta}(z)$ is a $4\times4$ matrix of coefficients and $\hat{\xi}=(\hat{X}_{A},\hat{Y}_{A},\hat{X}_{B},\hat{Y}_{B})^T$. The formal solution of this equation is given by $\hat{\xi}(z)=\mathbf{S}(z)\, \hat{\xi}(0)$, with $\mathbf{S}(z)=\exp\{\int_{0}^{z}\mathbf{\Delta}(z')\, d z' \}$. This is a linear unitary operator which contains the full evolution of our quantum system. This propagation matrix $\mathbf{S}(z)$ is given by the following eight independent coefficients as
\begin{align} \label{Sgen}
\nonumber
S_{1,1}=S_{3,3}\equiv \frac{1}{2}\{ (C_{K_{-}}+C_{K_{+}}  ) - (\Gamma_{-} S_{K_{-}} + \Gamma_{+} S_{K_{+}} ) S_{\phi} \}, \\   \nonumber
S_{1,2}=S_{3,4}\equiv \frac{1}{2}\{ (\Gamma_{-} C_{\phi} +\Lambda_{-}) S_{K_{-}}+(\Gamma_{+} C_{\phi} +\Lambda_{+}) S_{K_{+}}  \}, \\ \nonumber
S_{1,3}=S_{3,1}\equiv \frac{1}{2}\{ (C_{K_{-}}-C_{K_{+}}  ) - (\Gamma_{-} S_{K_{-}} - \Gamma_{+} S_{K_{+}} ) S_{\phi} \}, \\ \nonumber
S_{1,4}=S_{3,2}\equiv \frac{1}{2}\{ (\Gamma_{-} C_{\phi} +\Lambda_{-}) S_{K_{-}}-(\Gamma_{+} C_{\phi} +\Lambda_{+}) S_{K_{+}}  \}, \\ \nonumber
S_{2,1}=S_{4,3}\equiv \frac{1}{2}\{ (\Gamma_{-} C_{\phi} -\Lambda_{-}) S_{K_{-}}+(\Gamma_{+} C_{\phi} -\Lambda_{+}) S_{K_{+}} \}, \\ \nonumber
S_{2,2}=S_{4,4}\equiv \frac{1}{2}\{ (C_{K_{-}}+C_{K_{+}}  ) + (\Gamma_{-} S_{K_{-}} + \Gamma_{+} S_{K_{+}} ) S_{\phi} \}, \\ \nonumber
S_{2,3}=S_{4,1}\equiv \frac{1}{2}\{ (\Gamma_{-} C_{\phi} - \Lambda_{-}) S_{K_{-}}-(\Gamma_{+} C_{\phi} -\Lambda_{+}) S_{K_{+}}  \}, \\ 
S_{2,4}=S_{4,2}\equiv \frac{1}{2}\{ (C_{K_{-}}-C_{K_{+}}  ) + (\Gamma_{-} S_{K_{-}} - \Gamma_{+} S_{K_{+}} ) S_{\phi} \},
\end{align}
where we have defined the effective coupling $K_{\pm}=\sqrt{((\Delta\beta/2) \pm C)^{2}-4 \eta^{2}}$ and the dimensionless variables $C_{K_{\pm}}\equiv \cos(K_{\pm} z )$, $S_{K_{\pm}} \equiv \sin(K_{\pm} z)$, $C_{\phi}\equiv \cos(\phi)$, $S_{\phi} \equiv \sin(\phi)$, $\Gamma_{\pm}\equiv 2\eta/K_{\pm}$ and $\Lambda_{\pm}\equiv C/K_{\pm}$. From Equations (\ref{Sgen}) we can easily calculate the mean number of generated signal photons in each waveguide, given by
\begin{equation}\label{Ns}
N_{s}=2\eta^{2}(\frac{S^{2}_{K_{+}}}{K^{2}_{+}}+\frac{S^{2}_{K_{-}}}{K^{2}_{-}}).
\end{equation}
Note that, in the case of PPM ($\Delta\beta=0$) 
\begin{equation}\label{NsPPM}
N_{s}=4\eta^{2}(S_{K}/K)^{2},
\end{equation}
with an unique effective coupling $K_{\pm}\equiv K=\sqrt{C^{2}-4\eta^{2}}$, and in the case of no coupling ($C=\Delta\beta=0$), the usual  
\begin{equation}\label{NsS}
N_{s}=\sinh^{2}(2\eta z)
\end{equation}
 is recovered \cite{Barral2017}. 

We base our analysis of C-QPM on the comparison of the ouputs of the NDC in the PPM and C-QPM regimes. We also compute and display the relevant $\Delta\beta$-QPM in a single waveguide for comparison.

\begin{figure}[t]
  \centering
    \subfigure{\includegraphics[width=0.43\textwidth]{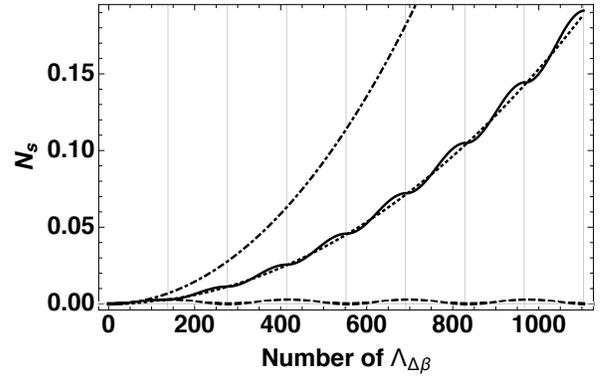}}
\vspace {0cm}\,
\hspace{0cm}\caption{\label{F2}\small{Mean number of signal photons in waveguide {\it a} (or {\it b}) versus propagation distance in terms of the number of phase mismatch periods $\Lambda_{\Delta\beta}$ : for PPM with $\eta\equiv\eta_{\Delta\beta}=(2/\pi)\, \eta_{0}$, without coupling (\NBc{dot-dashed}) and with coupling (\NBc{dashed}), for C-QPM with a pump input phase $\phi=-\pi/2$ and $\eta\equiv\eta_{0}$  (\NBc{solid}). PPM with $\eta\equiv\eta_{C}=(2/\pi)^{2}\, \eta_{0}$ and no coupling (\NBc{dotted}) is also shown for comparison. $\eta_{0}=15.10^{-3}$ mm$^{-1}$, $C=36.10^{-2}$ mm$^{-1}$ and $\Delta\beta=20.10^{-2}\, \mu m^{-1}$. The vertical lines show the phase-reversal lengths $L_{p}= 139\times 2 L_{c} = 4.37$ mm.}}
\end {figure}

\begin{figure}[p]
  \centering
    \subfigure{\includegraphics[width=0.425\textwidth]{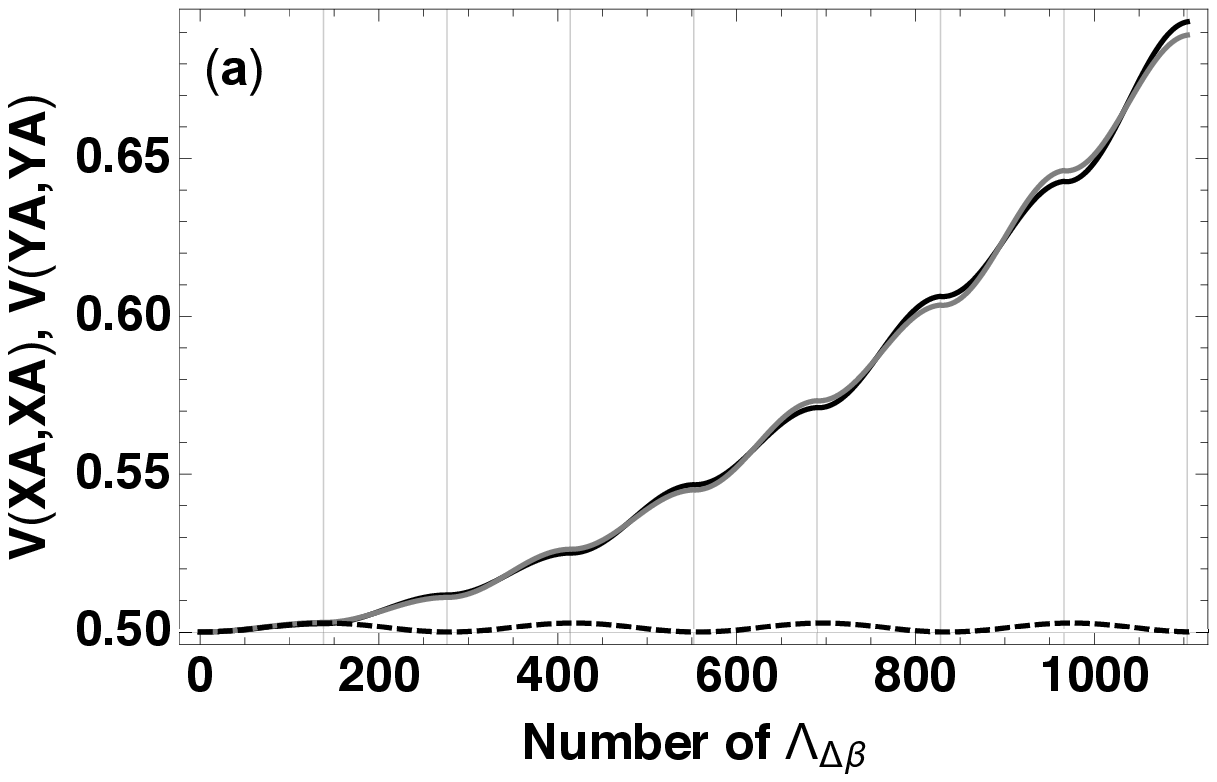}}
    \vspace {-0.2cm}
    \subfigure{\includegraphics[width=0.42\textwidth]{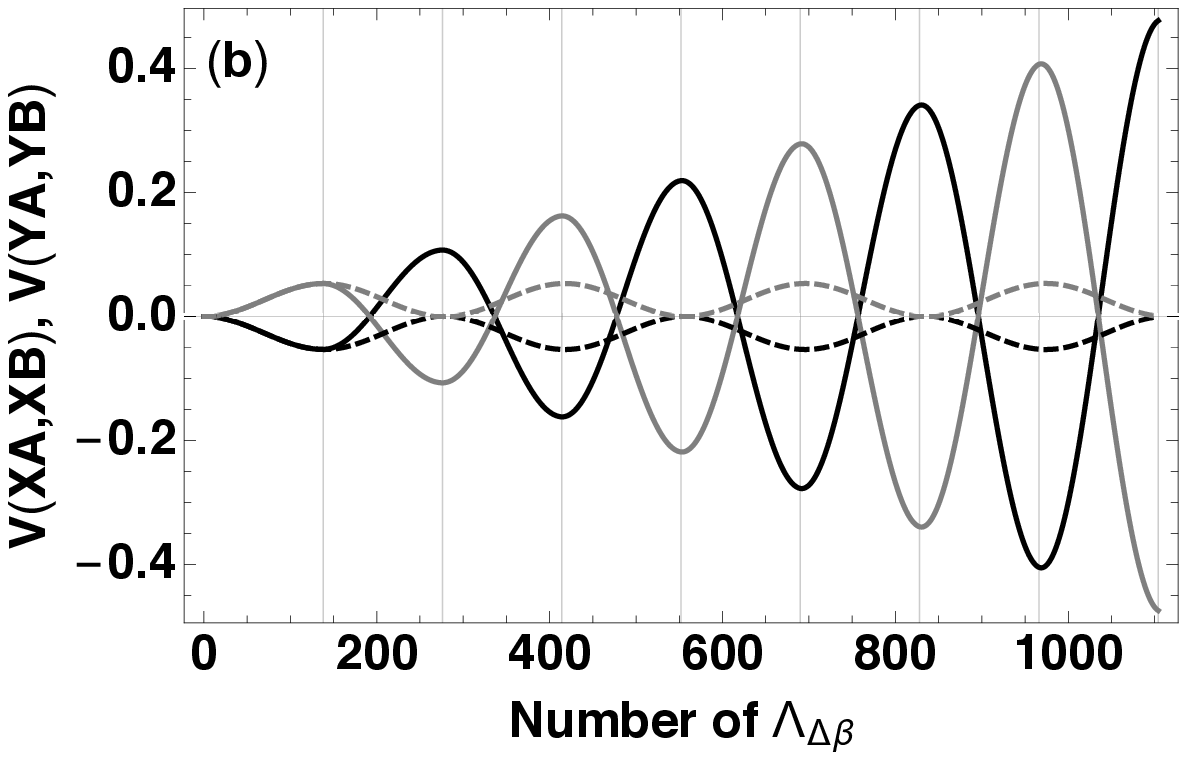}}
    \vspace {-0.2cm}
     \subfigure{\includegraphics[width=0.42\textwidth]{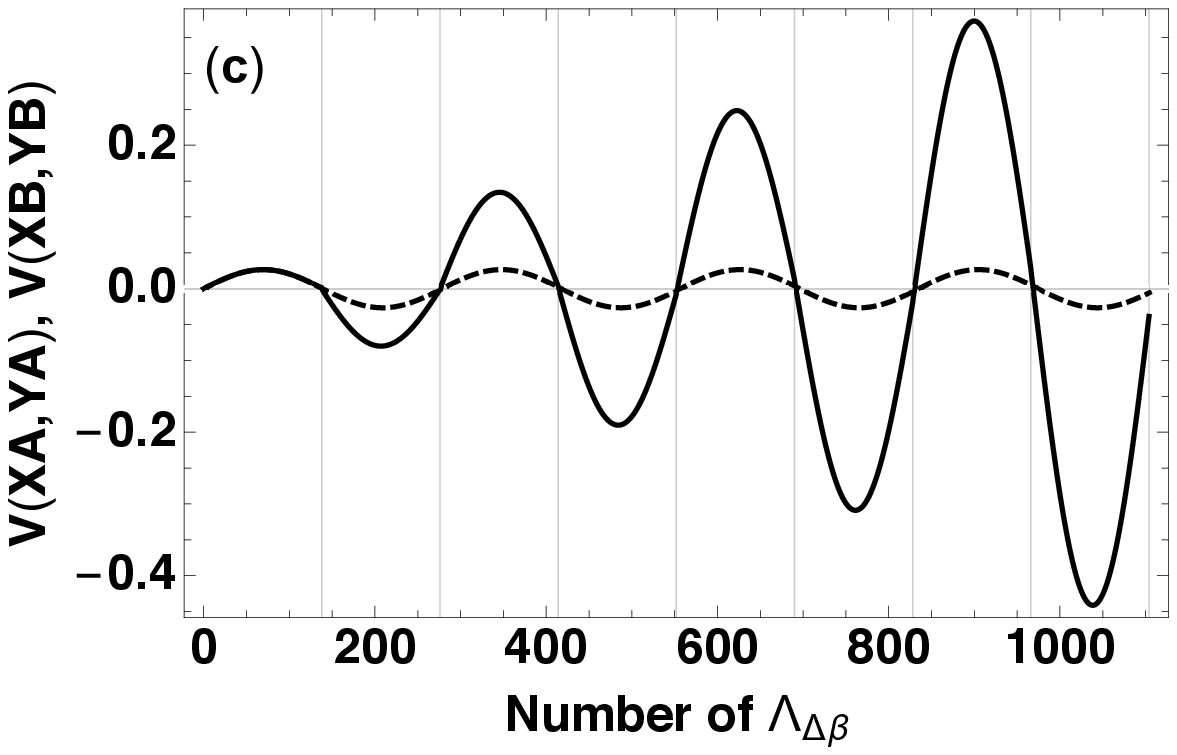}}
     \vspace {-0.2cm}
     \subfigure{\includegraphics[width=0.42\textwidth]{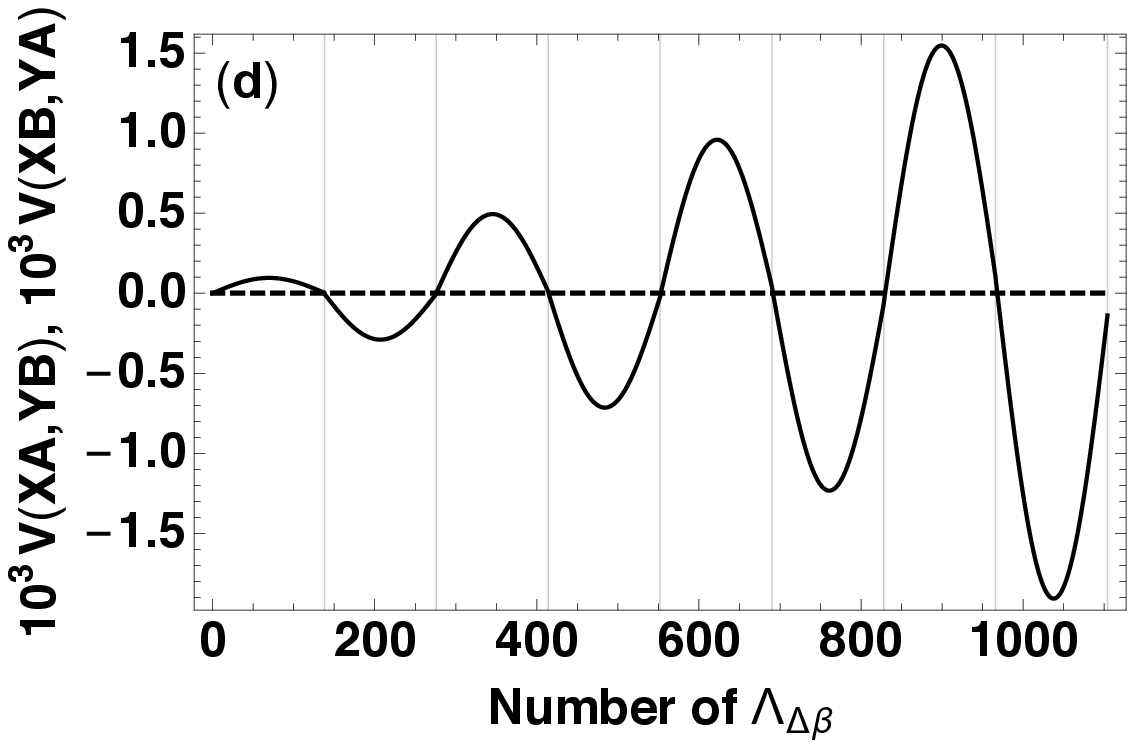}}
\vspace {0cm}\,
\hspace{0cm}\caption{\label{F3}\small{Elements of the covariance matrix for PPM with $\phi=0$ and $\eta\equiv\eta_{\Delta\beta}$ (\NBc{dashed}), and C-QPM with $\phi=-\pi/2$ and $\eta\equiv\eta_{0}$ (\NBc{solid}) versus propagation distance in terms of the number of phase mismatch periods $\Lambda_{\Delta\beta}$. From top to bottom: (a) variance of the amplitude and phase quadratures: $V(X_{A}, X_{A})$ (\NBc{black}) and $V(Y_{A}, Y_{A})$ (\NBc{gray}), respectively; (b) correlation elements $V(X_{A}, X_{B})$ (\NBc{black}) and $V(Y_{A}, Y_{B})$ (\NBc{gray}), (c) $V(X_{A}, Y_{A})$ and $V(X_{B}, Y_{B})$, and (d) $10^{3}\, V(X_{A}, Y_{B})$ and $10^{3}\, V(X_{B}, Y_{A})$. $\eta_{0}=15.10^{-3}$ mm$^{-1}$, $C=36.10^{-2}$ mm$^{-1}$ and $\Delta\beta=20.10^{-2}\, \mu m^{-1}$. The vertical lines show the phase-reversal lengths $L_{p}= 139\times 2 L_{c} = 4.37$ mm.}}
\end{figure}

\NB{It is common to consider PPM to be fulfilled in theoretical works about nonlinear waveguide arrays \cite{Herec2003}. This model is exact in the case of birefringence-based phase matching}. Then $\Delta \beta=0$ and the non-zero coefficients in the evolution matrix given by Equations (\ref{Sgen}) are 
\begin{align}\label{Spart}
\nonumber
S_{1,1}=S_{3,3}&= C_{K} - \Gamma  S_{K} S_{\phi}, \\ \nonumber
S_{2,2}=S_{4,4}&=C_{K} + \Gamma S_{K} S_{\phi}, \\   \nonumber
S_{1,2}=S_{3,4}&=S_{2,1}=S_{4,3}=\Gamma C_{\phi} S_{K}, \\ \nonumber
S_{1,4}=S_{3,4}&=-S_{2,3}=-S_{4,1}=-\Lambda S_{K}. \nonumber
\end{align}
These solutions have been recently used to study the continuous-variables capabilities of this device \cite{Herec2003, Barral2017}. \NB{The intensity in Equation (\ref{NsPPM}), the noise squeezing and the entanglement oscillate} with $z$ with maximum values at beat lengths $(2l+1)L_{b}$ with $L_{b}=\pi/2K$ and $l$ a positive integer (Figures \ref{F2}, \ref{F3}a and \ref{F4}a (dashed) )\cite{Herec2003, Barral2017}. This is a widely used model of NDC. For a single waveguide, PPM and $\Delta\beta$-QPM provide the same results in terms of gain and noise squeezing apart from a reduction in the effective nonlinearity, $\eta\equiv\eta_{\Delta\beta}=(2/\pi) \eta_{0}$ for first order $\Delta\beta$-QPM. However, as outlined above, there is also a phase retardation effect that has to be borne in mind in the case of two evanescently coupled waveguides. Thus, we discuss below the realistic case of $\Delta \beta$-QPM in the NDC.


Now, let us make a reminder of the $\Delta \beta$-QPM technique and see what happens when applied to coupled waveguides. The relative phase between the nonlinear polarization and the pump beam varies linearly along propagation through the nonlinear material. This phase drives a periodic cascade effect from downconversion to upconversion, which periodically wipes away the generated signal light \cite{Armstrong1962}. For a single waveguide, the generated signal intensity oscillates as a function of the propagation distance with a typical period $\Lambda_{\Delta\beta}=\pi/\sqrt{(\Delta\beta/2)^{2}-4 \eta^{2}}\equiv2L_{c}$, with $L_{c}$ the coherence length (see Equation (\ref{Ns}) for $C=0$). Usually $\Delta\beta \gg \eta$, and the period is approximately equal to $2\pi/\Delta\beta$. The $\Delta \beta$-QPM technique consists in inserting domains with an inverted nonlinear polarization, changing $\eta$ into $- \eta$ in the second half coherence length in order to build up the nonlinear interaction, achieving large parametric gains. Therefore, after the first complete period $\Lambda_{\Delta\beta}$, the propagation matrix Eq. (\ref{Sgen}) is
\begin{equation} \nonumber
\mathbf{S}_{\Delta\beta}(2 L_{c},\eta \rightarrow -\eta, 0)=\mathbf{S}(L_{c}, -\eta, 0) \mathbf{S}(L_{c}, \eta, 0),
\end{equation}
where $\pm\eta \rightarrow \mp\eta$ stands for an inversion of the nonlinear polarization and {$\mathbf{S}(z, \eta, C)$ for the propagator $\mathbf{S}$ at given length, nonlinearity and coupling}. After propagation through {\it n} periods, the propagation matrix for $\Delta\beta$-QPM in a single waveguide is
\begin{equation}\label{Sdb}
\mathbf{S}_{\Delta\beta}(2 n L_{c},\eta \rightarrow -\eta, 0)=(\mathbf{S}_{\Delta\beta}(2 L_{c},\eta \rightarrow -\eta, 0))^{n}.
\end{equation}
This equation is easily computed if $\mathbf{S}_{\Delta\beta}(2 L_{c},\eta \rightarrow -\eta, 0)$ presents independent eigenvectors $\mathbf{u}$. In this case, $(\mathbf{S}_{\Delta\beta}(2 L_{c},\eta \rightarrow -\eta, 0))^{n}= \mathbf{u} \mathbf{\lambda}^{n} \mathbf{u}^{-1}$, with $\mathbf{\lambda}$ the diagonal matrix of eigenvalues of $\mathbf{S}_{\Delta\beta}(2 L_{c},\eta \rightarrow -\eta, 0)$. Using this approach, analytical solutions to Equation (\ref{Sdb}) have been found \cite{Bencheikh1995}.

In the case of evanescently coupled waveguides there is a second cause of phase mismatch: the coupling $C$. Since in general  $\Delta\beta \gg C$, the fast oscillation period is approximately $2\pi/\Delta\beta$, thus the $\Delta\beta$-QPM uniform grating can be safely set as $\Lambda_{\Delta\beta}= 2\pi/\Delta\beta$. Similarly to the case of the PPM NDC, the intensity in the coupler oscillates with maxima obtained at beat lengths $(2l+1)L_{b}$. The physics in NDC is the same as for wavevector mismatch in a single waveguide, but the oscillation in NDC is in general much slower as $L_{b} \gg L_{c}$. Thus following the $\Delta\beta$-QPM strategy above, we propose to invert super-domains with a period $2 L_{b}$ to compensate for the phase retardation generated by the coupling. However, the beat length is pump-power dependent $L_{b}\equiv L_{b}(\eta)$. For typical input powers in the continuous wave regime --tens or hundreds of mW-- $C>20\eta$, such that $L_{b}\approx\pi/2C\equiv L_{p}$. We use then $P=100$ mW as input power in the remainder of the paper. The power-independent linear-coupling beat length --or phase-reversal length $L_{p}$-- can then be safely used as a design parameter. The superperiod is thus given by $\Lambda_{C}=2 L_{p}$ (phase-reversal grating, Figure \ref{F1}). Note that when $C\lessapprox 2\eta$, the solutions are no longer oscillatory and the mismatch rephasing is not necessary \cite{Herec2003}. However, this regime is far from being technologically accessible in CW traveling-wave integrated devices.

For the sake of simplicity, let us choose the experimental phase-reversal length as an even number of coherence lengths $L_{p}=2 n L_{c}$. This fixes the value of the coupling constant with respect to the wavevector mismatch to $C=\Delta\beta/4n$. After the first complete superperiod, the propagation matrix $\mathbf{S}(z)$ in the C-QPM NDC is:
\begin{equation} \label{Sdc}
\mathbf{S}_{C}(2 L_{p})=\mathbf{S}_{\Delta\beta}(2 n L_{c}, -\eta \rightarrow \eta, C) \mathbf{S}_{\Delta\beta}(2 n L_{c}, \eta \rightarrow -\eta, C),  \nonumber
\end{equation}
where we have used Eq. (\ref{Sdb}) and include in addition the effect of the coupling $C$. Again, after a number $m$ of superperiods we have
\begin{equation} \label{Sdd}
\mathbf{S}_{C}(2 m L_{p})=(\mathbf{S}_{C}(2 L_{p}))^{m}.
\end{equation}
This matrix contains the full evolution of any classical or quantum state of light propagating in the device. As above, we can make use of the matrix diagonalization of $\mathbf{S}_{C}(2 L_{p})$ to calculate numerically the result of Equation (\ref{Sdd}) \cite{Note1}. 

The signal light intensity generated in each waveguide can be readily extracted from this evolution operator generalizing Equation (\ref{Ns}). Figure \ref{F2} shows the mean number of signal photons in waveguide {\it a} ({\it b}) for the different cases of phase matching we introduced : PPM and C-QPM. We use the relation in Equation (\ref{RelEta}) for a fair comparison between cases. We give PPM without coupling as given in Equation (\ref{NsS}) (\NBc{dot-dashed}), PPM with a coupling $K$ as given in Equation (\ref{NsPPM}) with $\eta\equiv\eta_{\Delta\beta}$ (\NBc{dashed}),  and C-QPM with $\phi=-\pi/2$ and $\eta\equiv\eta_{0}$ as calculated from Equation (\ref{Sdd}) (\NBc{solid}). {A similar figure is obtained for $\Delta\beta$-QPM after a scale change from $L_{p}$ to $2 L_{c}$. This type of figure appears ubiquitously in the $\Delta\beta$-QPM literature \cite{Fejer1992}. We demonstrate thus how the C-QPM similarly rephases the coupling-based phase mismatch. For a further comparison with C-QPM, we also display the intensity obtained via PPM with $\eta\equiv\eta_{C}$ and no coupling (Equation (\ref{NsS})) (\NBc{dotted}), which is the same we can retrieve via the simplified model and the supermode Equations (\ref{QS2}). This establishes C-QPM as a first-order coupling QPM for the directional coupler supermodes producing a clear intensity buildup of the signal in a NDC in the SPDC regime.}

\section{Noise squeezing, CV entanglement and EPR steering} The most interesting observables of the NDC in terms of squeezing and CV entanglement are the second-order moments of the quadrature operators, properly arranged in the covariance matrix $\mathbf{V}$ \cite{Adesso2014}. The elements of this matrix can be efficiently measured by means of homodyne detection \cite{Dauria2009}. The covariance matrix at any propagation plane $z$ is given by $\mathbf{V}(z)=\mathbf{S}(z)\, \mathbf{V}(0) \,\mathbf{S}^{T}(z)$, where $\mathbf{V}(0)=(1/2)\, \mathbf{1}$ is the covariance matrix related to the vacuum state of the input signal modes, with 1/2 the shot noise. Evolution of squeezing $V_{ii}=V(\xi_{i}, \xi_{i})$ and quantum correlations $V_{ij}=V(\xi_{i}, \xi_{j})$ can be obtained at any length $z$ from the elements of this matrix. 

Figure \ref{F3} shows all the relevant elements of the covariance matrix for a realistic implementation of C-QPM. We compare PPM with $\phi=0$ and $\eta\equiv\eta_{\Delta\beta}$ (\NBc{dashed}) with C-QPM with $\phi=-\pi/2$ and $\eta\equiv\eta_{0}$ (\NBc{solid}) due to the $\Delta \beta$-QPM-based phase retardation \cite{Bencheikh1995}. The vertical lines correspond to the selected experimental lengths (phase-reversal lengths $L_{p}$) where the superperiods are inverted. Figure \ref{F3}a shows the amplitude and phase squeezing of the signal mode A in waveguide {\it a}. The same result is obtained for mode B due to the symmetry of $\mathbf{V}$. \NB{Quantum noise buildup above shot noise is clearly displayed. This is a signature of entanglement: the fields-superposition quantum noise, i.e. the supermode quadratures noise, is squeezed whereas the individual-fields quantum noise is super-Poissonian \cite{Ou1992}.} As expected, after few phase-reversal lengths, amplitude and phase squeezing have different evolutions in the C-QPM case, in contrast with the PPM case where they overlap. This is due to the evolving coupling-based phase generated through propagation in $\Delta \beta$-QPM structures which retards the phase quadrature with respect to the amplitude quadrature. Figures \ref{F3}b, c and d show the buildup evolution of all correlations in the device. These elements follow a similar evolution for QPM and PPM except for the {elements $V(X_{A}, Y_{B})$ ($V(X_{B}, Y_{A})$) in Figure \ref{F3}d where the ordinate axis has been expanded by a factor of $10^{3}$.} A small decoupling between the phase-reversal lengths $L_{p}$ and the coupling beat lengths $L_{b}$ is observed for all the elements of $\mathbf{V}$ after a large number of superperiods. This effect is due to the nonlinear correction to the coupling beat length implicit in Equation (\ref{NsPPM}). However, for typical cw pump powers and PPLN lengths this shift is negligible.

The input-output transformation generated by a NDC in the undepleted regime, can be decomposed {into elementary transformations} generated by a beam splitter and optical parametric amplifiers with suitable parameters \cite{Fiurasek2000}. Using this substituting scheme, we identify the quantum states $\vert \psi (z)\rangle$ generated at each propagation plane $z$ from the off-diagonal elements $V(\xi_{i}, \xi_{j})$ of the covariance matrix $\mathbf{V}$. At lengths $z_\text{EPR}(m)=m L_{b}$, the quantum state is a two-mode squeezed (EPR) state given by $|\psi (z_{EPR})\rangle=\hat{S}_{A B}[r(z_{EPR})] \vert 0_{A}\, 0_{B} \rangle$ with \NBc{$\hat{S}_{A B}[r]=\exp{\{-r[\hat{A}^{\dagger}\,\hat{B}^{\dagger}-\hat{A}\,\hat{B}]\}}$} and where $V(X_{A}, X_{B})=-V(Y_{A}, Y_{B})\neq0$; whereas at lengths $z_\text{NOON}(m)=(2m+1)L_{b}/2$ the quantum state is a \NBc{separable} two single-mode squeezed state given by $|\psi (z_{NOON})\rangle=\hat{S}_{A}[r(z_{NOON})] \hat{S}_{B}[r(z_{NOON})] \vert 0_{A}\, 0_{B} \rangle$ with \NBc{$\hat{S}_{A}[r]=\exp{\{-(r/2)[\hat{A}^{\dagger\,2}-\hat{A}^{2}]\}}$}, and where $V(X_{A}, Y_{A})=V(X_{B}, Y_{B})\neq 0$. \NBc{In fact, the state $|\psi (z_{NOON})\rangle$ is not perfectly separable due to the small contribution of the cross-correlation elements $V(X_{A}, Y_{B})=V(X_{B}, Y_{A})$. These elements are 100 times lower than the auto-correlation elements $V(X_{A}, Y_{A})=V(X_{B}, Y_{B})$, thus making the state separable in practical terms. At lengths  $z_\text{NOON}(m)$ the state can then be written as $|\psi (z_{NOON})\rangle\approx \vert 0_{A} 0_{B} \rangle -(r/\sqrt{2})(\vert 2_{A} 0_{B} \rangle + \vert 0_{A} 2_{B} \rangle) + O(r^2)$. Thus, the device generates two-photon NOON states with probability $r^2/2$ when low pump power is used in order to avoid the contribution of higher number-of-photon states.} In contrast to the PPM NDC case where the squeezing parameter is bounded, here the squeezing parameter $r(z)$ is building up with propagation. The squeezing parameter can be extracted from the covariance matrix values through a Bloch-Messiah decomposition \cite{Arzani2018}.

\begin{figure}[t]
  \centering
    \subfigure{\includegraphics[width=0.43\textwidth]{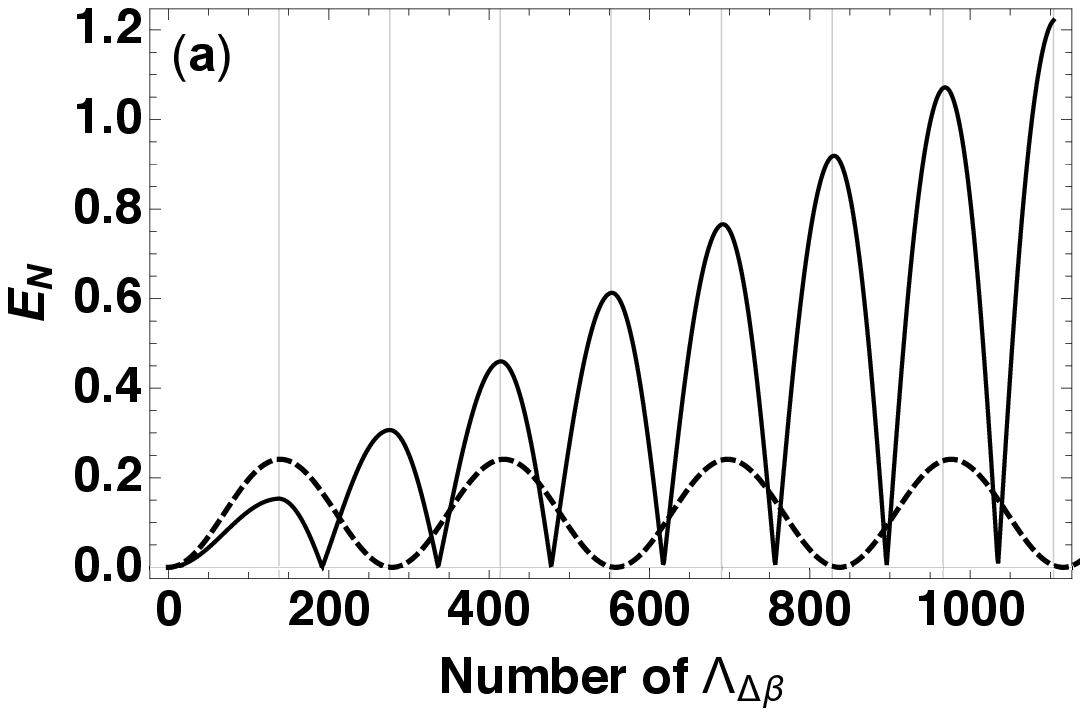}}
    \vspace {-0.2cm}
    \subfigure{\includegraphics[width=0.43\textwidth]{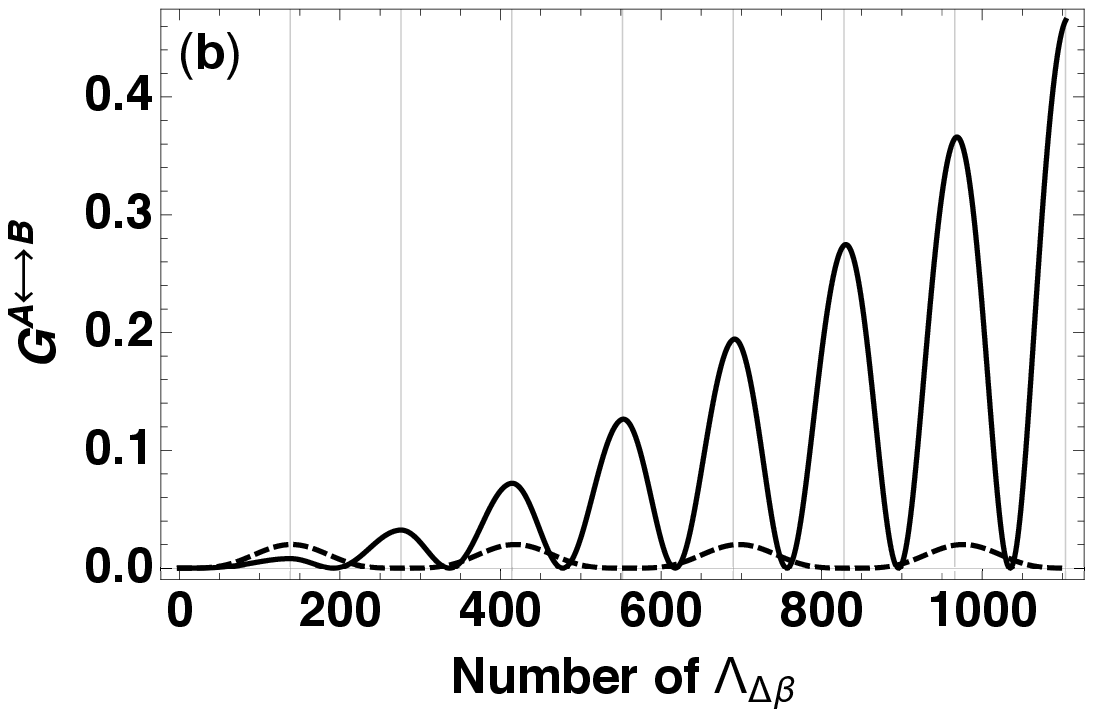}}
\vspace {0cm}\,
\hspace{0cm}\caption{\label{F4}\small{Logarithmic negativity (a) and Gaussian steering (b) for PPM with $\phi=0$ and $\eta\equiv\eta_{\Delta\beta}$ (\NBc{dashed}), and C-QPM with $\phi=-\pi/2$ and $\eta\equiv\eta_{0}$ (\NBc{solid}) versus propagation distance in terms of the number of phase mismatch periods $\Lambda_{\Delta\beta}$. $\eta_{0}=15.10^{-3}$ mm$^{-1}$, $C=36.10^{-2}$ mm$^{-1}$ and $\Delta\beta=20.10^{-2}\, \mu m^{-1}$. The vertical lines show the phase-reversal lengths $L_{p}= 139\times 2 L_{c} = 4.37$ mm.}}
\end {figure}

The amount of CV entanglement of the two-mode system is easily quantified through the logarithmic negativity $E_{\mathcal{N}}$ \cite{Vidal2002}. This entanglement witness is based on the {Peres-Horodecki-Simon} criterion, which establishes that a quantum state is entangled if the partially transposed density matrix is non-positive. {$E_{\mathcal{N}}$} can be obtained from the covariance matrix $\mathbf{V}$ and is defined in such a way that any value $E_{\mathcal{N}}>0$ indicates entanglement. For two-mode Gaussian pure systems, it reads
\begin{equation}\label{EN}
\NBc{E_{\mathcal{N}}=\max\{0,- 2 \log_{2} (\sqrt{\frac{1}{2\mu_{A}} - \frac{1}{2}} - \sqrt{\frac{1}{2\mu_{A}} + \frac{1}{2} } )\},}
\end{equation}
with $\mu_{A}$ the partial purity related to mode A, $\mu_{A}= [4 (V(X_{A}, X_{A}) V(Y_{A}, Y_{A}) - V(X_{A}, Y_{A})^{2})]^{-1/2}$ \cite{Vidal2002}.  Figure \ref{F4}a displays the buildup evolution of the logarithmic negativity.  $E_{\mathcal{N}}$ linearly increases in the {$z_\text{EPR}(m)$} planes with the number of superperiods $m$, leading to strongly entangled two-mode squeezed states. For instance, for $z=7 L_{b}\approx$ 3.1 cm, a typical length in PPLN waveguides, we have {thus designed} a buildup of $\approx7\times(2/\pi)=4.5$ times with respect to the PPM case, since $E_{\mathcal{N}}^{CQPM}(\eta,L_{b})\approx (2/\pi)E_{\mathcal{N}}^{PPM}(\eta,L_{b})$ for $\eta$ small.



{An even stronger} type of quantum correlation is EPR steering. This quantum feature allows one party, Alice, to change the state of a distant party, Bob, by exploiting their shared entanglement \cite{Wiseman2007}. From a quantum information processing  perspective, quantum steering corresponds to the task of verifiable entanglement distribution by an untrusted party, and it has been shown that this quantum protocol provides security in one-sided device-independent quantum key distribution \cite{Branciard2012}. Recently, a quantum steering witness for Gaussian systems has been proposed \cite{Kogias2015}. In the case under study, a two-mode pure system, the Gaussian $A \rightarrow B$ steerability is given by \cite{Note2}
\begin{equation}\label{GAB}
\NBc{\mathcal{G}^{A \rightarrow B}=\max \{0, -\log_{2} \mu_{A} \}.}
\end{equation}
Since we are dealing with pure Gaussian states, thus symmetric, the steering is also symmetric with $\mathcal{G}^{A \rightarrow B}=\mathcal{G}^{B \rightarrow A}\equiv \mathcal{G}^{A \leftrightarrow B}$.  Figure \ref{F4}b shows the buildup evolution of the Gaussian steering. $\mathcal{G}^{A \leftrightarrow B}$ increases faster than $E_{\mathcal{N}}$ at {$z_\text{EPR}(m)$} distances. \NBc{This feature is the consequence of a faster-than-linear decrease of the partial purity $\mu_{A}(z)$ at $z_{\text{EPR}}(m)$ lengths (not shown) and the definitions of $E_{\mathcal{N}}(\mu_{A})$ and $\mathcal{G}^{A \leftrightarrow B}(\mu_{A})$, Equations (\ref{EN}) and (\ref{GAB}) respectively.} Thus, as Gaussian steering is a measure of how useful entanglement is in certain quantum protocols, the C-QPM NDC represents a potential candidate as resource for CV quantum information processing. \NBc{To our knowledge, this is the first analysis of Gaussian steering in a realistic integrated system.}

\section{Discussion and conclusion}
\begin{figure}[t]
\centering
\includegraphics[width=0.43\textwidth]{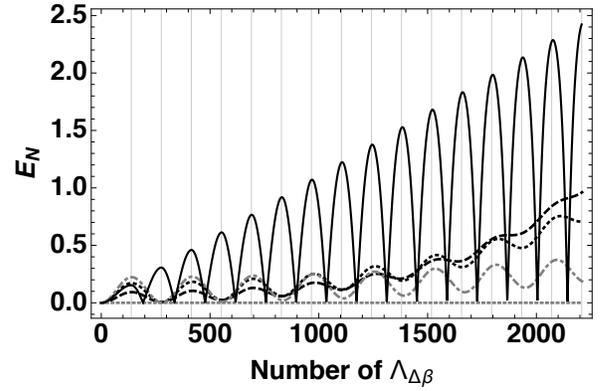}
\vspace {0cm}\,
\hspace{0cm}\caption{\label{F5}\small{Logarithmic negativity for PPM with \NBc{$\phi_{a}=0$} and \NBc{$\Delta\phi\equiv\phi_{a}-\phi_{b}$}$=\pi/4$ (\NBc{dot-dashed, gray}),  $\Delta\phi=\pi/2$ (\NBc{dotted, black}), $\Delta\phi=3\pi/4$ (\NBc{dashed, black}) and $\Delta\phi=\pi$ (\NBc{small dashing, gray}), and C-QPM with $\phi_{a}=-\pi/2$ and $\Delta\phi=0$ (\NBc{solid, black}) versus propagation distance in terms of the number of phase mismatch periods $\Lambda_{\Delta\beta}$. $\eta\equiv\eta_{0}=15.10^{-3}$ mm$^{-1}$, $C=36.10^{-2}$ mm$^{-1}$ and $\Delta\beta=20.10^{-2}\, \mu m^{-1}$. The vertical lines show the phase-reversal lengths $L_{p}= 139\times 2 L_{c} = 4.37$ mm.}}
\end {figure}
We finally analyze the robustness of the C-QPM approach for efficient generation of squeezed light. In practice, the {C-QPM approach} does not increase the propagation losses with respect to $\Delta\beta$-QPM PPLN waveguides. A state-of-the-art value for signal field losses is $\gamma_{s}\approx 0.14$ dB cm$^{-1}$ in reverse-exchange PPLN waveguides \cite{Lenzini2018}. {The influence of linear losses, like scattering or absorption,} on the fields quadratures can be easily included in our analysis by inserting in Equations (\ref{Sgen}) fictitious beam splitters with effective transmittivities proportional to the losses corresponding to each mode \cite{Barral2018}. Our simulations  point out that the noise squeezing and entanglement buildup produced by C-QPM is quite robust under these values of losses. {For instance, a fall lower than 2\% is obtained from an ideal 3 dB squeezing using the above value of losses}. In the case of imperfect C-QPM due to a small mismatch in the superperiods the entanglement {still increases although with reduced efficiency}, as it happens for $\Delta \beta$-QPM in single waveguides. However, typical $\Lambda_{C}$ in PPLN waveguide arrays are {of} the order of {mm} or {cm}, thus easily realizable with the current state-of-the-art technology \cite{Alibart2016}. As above commented, since the coupling length $L_{b}$ depends on the input pump power, there will be always an small shift between the phase-reversal length $L_{p}$ and $L_{b}$. This mismatch grows with the number of superperiods. However the entanglement buildup is not affected to a great extent as shown in Figure \ref{F5} (\NBc{solid, black}). We have also compared the performance of C-QPM with that obtained engineering the pumps phase, i.e. using $\Delta\phi\equiv\phi_{a}-\phi_{b}\neq0$ \cite{Herec2003, Note0}. Figure \ref{F5} compares the amount of entanglement generated in C-QPM (\NBc{solid, black}) with that obtained by only tuning of the pumps phase difference $\Delta\phi=\pi/4$ (\NBc{dot-dashed, gray}), $\pi/2$ (\NBc{dotted, black}), $3\pi/4$ (\NBc{dashed, black}) and $\pi$ (\NBc{small dashing, gray}). In contrast with previous figures, here we compare directly both cases, i.e. using $\eta=\eta_{0}$ for both. We have found that for the parameters used along the paper, C-QPM is more efficient in terms of entanglement than any $\Delta\phi\neq0$ up to $z=25 L_{b}\approx 11$ cm (not shown). To our knowledge, the largest lithium niobate chips are $\approx 7$ cm long, but typical PPLN waveguides cover the 2-4 cm range, to avoid problems related to inhomogeneities. To find a device length where $\Delta\phi\neq0$ beats C-QPM available with present technology, such as $z=7 L_{b}\approx 3.1$ cm, $P\geq625$ mW per waveguide is necessary. Note that for long distances and/or high input power, the undepleted pump approximation is no longer valid. In such case the depletion of the pump has to be taken into account \cite{Barral2017, Barral2018}. Thus we conclude that C-QPM is the best option available as of today in terms of compactness and efficiency to generate quantum correlations in {cw}-PPLN directional couplers. 

The analysis here presented can be extended to any device supporting supermodes. Particularly, bipartite and multipartite entanglement enhancement can be obtained in nonlinear waveguide arrays in the SPDC regime under {well-chosen} pump configurations and suitable phase reversal periods.
A detailed analysis will be presented elsewhere. Finally, note that this scheme also applies to second harmonic generation, where entanglement of the generated signal fields has been predicted \cite{Barral2018}.


\section*{Acknowledgements} This work was supported by the Agence Nationale de la Recherche through the INQCA project (grant agreement number PN-II-ID-JRP-RO-FR-2014-0013), the Paris Ile-de-France region in the framework of DIM SIRTEQ through the project ENCORE, and the Investissements d'Avenir program (Labex NanoSaclay, reference ANR-10-LABX-0035).

\end{document}